# Next Generation Crowdsourcing for Collective Intelligence

JOHN PRPIĆ – Faculty of Business Administration, Technology and Social Sciences, Lulea University of Technology

---

## 1: INTRODUCTION

New techniques leveraging IT-mediated crowds such as Crowdsensing, Situated Crowdsourcing, Spatial Crowdsourcing, and Wearables Crowdsourcing have now materially emerged. These techniques, here termed next generation Crowdsourcing, serve to extend Crowdsourcing efforts beyond the heretofore dominant desktop computing paradigm. Employing new configurations of hardware, software, and people, these techniques represent new forms of organization for IT-mediated crowds (Prpić et al 2015; 2015b, 2015c, 2015d, Prpić & Shukla 2013; 2014; 2016). However, it is not known how these new techniques change the processes and outcomes of IT-mediated crowds for Collective Intelligence purposes? The aim of this exploratory work is to begin to answer this question. The work ensues by outlining the relevant findings of the first generation Crowdsourcing paradigm, before reviewing the emerging literature pertaining to the new generation of Crowdsourcing techniques. Premised on this review, a collectively exhaustive and mutually exclusive typology is formed, organizing the next generation Crowdsourcing techniques along two salient dimensions common to all first generation Crowdsourcing techniques. As a result, this work situates the next generation Crowdsourcing techniques within the extant Crowdsourcing literature, and identifies new research avenues stemming directly from the analysis.

## 2: FIRST GENERATION CROWDSOURCING

Crowdsourcing, first named and described by Howe (2006), and then extended materially by Brabham, describes the use of IT to engage crowds for the purposes of completing tasks, solving problems or generating ideas. Differing from other established modes of organization, Crowdsourcing is a deliberate blend of bottom-up crowd-derived processes and inputs, combined with top-down goals set and initiated by an organization (Brabham 2008; 2013). As a problem solving, idea-generation, and production model, Crowdsourcing leverages the dispersed knowledge and energy of individuals in IT-mediated crowds (Brabham 2008; 2013, Hayek 1945, Prpić & Shukla 2013; 2014), through different configurations; such as micro-tasking at virtual labor markets (Ipeirotis & Paritosh 2011, Michelucci 2013), open collaboration through social media, wikis, and web-properties (Prpić et al 2015; 2015b) or tournament-based competitions at web platforms (Afuah & Tucci, 2012, Fayard et al 2014, Jeppesen & Lakhani 2010). To date, the vast majority of research and practice on Crowdsourcing is bound directly to the desktop computing paradigm (Goncalves et al 2015). However, with continued IT development, in concert with the continued diffusion of IT developments throughout our world, we are now beginning to observe new configurations of hardware, software, and people as IT-mediated crowds. In the next section, these new techniques are introduced and detailed in brief.

## 3: NEXT GENERATION CROWDSOURCING

Crowdsensing, Situated Crowdsourcing, Spatial Crowdsourcing, and Wearables Crowdsourcing represent new configurations of hardware, software, and people as IT-mediated crowds. These new techniques are discussed briefly in turn.

### 3.1 Crowdsensing

Crowdsensing (sometimes termed 'participatory sensing' or 'social sensing') leverages the built-in sensors in smartphone devices to gather environmental data such as location, temperature, and acceleration, as a result of human mobility, smartphone portability and wifi/mobile networking (Malatras & Beslay 2015, Sun et al 2015, Zenonos 2016). Crowdsensing techniques passively collect data through the smartphone sensor hardware, and autonomously supply the data content through wifi/mobile networks to the Crowdsourcing endeavor through time.

### 3.2 Situated Crowdsourcing

Situated Crowdsourcing is a technique that employs display hardware (for example tablets and other touchscreens) fixed to a particular geographic location, to generate inputs from human beings engaged at the geo-fenced terminal. Situated Crowdsourcing has focused on employing human skills at these fixed IT installations, therein, engaging human problem solving and idea generation skills (Goncalves et al 2013, Goncalves et al 2014, Goncalves et al 2014b, Goncalves et al 2014c, Heimerl et al 2012, Hosio et al 2014, Hosio et al 2015).

### 3.3 Spatial Crowdsourcing

Spatial Crowdsourcing uses multiple forms of IT (such as Mobile apps & Web platforms, Smartphone devices, and Wifi/Mobile networks) to animate individual human beings to perform specific actions in the physical environment (Cheng at al 2014, ul Hassan & Curry 2015, Cheng at al 2015, To et al 2016). Spatial Crowdsourcing can be found in a variety of contexts, including Emergency response (Goodchild & Glennon 2010), the Gig economy (Friedman 2014), and the Sharing economy (Cohen & Kietzmann 2014).

### 3.4 Wearables Crowdsourcing

Wearables leverage sensors built-into devices attached to the human body through apparel or accessories (Mann 1997, Pantelopoulos & Bourbakis 2010, Pascoe 1998, Wilde et al 2015). As the killer app of the 'quantified-self' movement (Swan 2009; 2012; 2013), wearables collect and transmit data only about the specific wearer of the device, and can only supply such individually-focused data to Crowdsourcing efforts incorporating them.

### 4: ANALYSIS

Figure 1 below depicts a typology of the next generation Crowdsourcing techniques predicated upon two dimensions; a) type of human participation and b) content-type, which are found to be salient to all first generation Crowdsourcing research and applications. The typology illustrates that Wearables Crowdsourcing and CrowdSensing generate inputs from IT-mediated crowds in a passive manner, without the need for the task-by-task involvement of the human participant. Conversely, it is illustrated that in Situated and Spatial Crowdsourcing, human intervention is necessary on a task-by-task basis, if inputs are to be derived from these IT-mediated crowds. Similarly, the typology illustrates that Wearables Crowdsourcing and Situated Crowdsourcing are solely focused upon intelligence from specific individuals, while CrowdSensing and Spatial Crowdsourcing are focused solely upon environmental intelligence.

**Figure 1 - Typology of Next Generation Crowdsourcing Techniques**

|  | Individual | Environmental |
|---|---|---|
| Passive | Wearables | CrowdSensing |
| Active | Situated Crowdsourcing | Spatial Crowdsourcing |

Type of Human Participation (vertical axis) / Type of Intelligence (horizontal axis)

**John Prpić - Next Generation Crowdsourcing for Collective Intelligence          Collective Intelligence 2016**

**5.0: DISCUSSION**

The research on human participation in Crowdsourcing has generally focused upon the intrinsic or extrinsic motivations of individuals forming, and participating in, IT-mediated crowds (Antikainen et al 2010, Antikainen & Vaataja 2010, Chandler & Kapelner 2013, Gerber & Hui 2013, Kaufmann & Schulze 2011, Majchrzak et al 2006, Miles & Mangold 2014, Morgan & Wang 2010, Raddick et al 2008, Rogstadius et al 2011, Rotman et al 2012, Yang & Cheng 2010). Due to the fact that these first generation Crowdsourcing techniques have been tied directly to the desktop computing paradigm, the research has always assumed (rightfully so) that human participation in an IT-mediated crowd is solely and completely active. In this conception of Crowdsourcing, an individual must always actively respond -ex post- to a certain stimulus supplied -ex ante- by the Crowdsourcing IT, or by other members of the IT-mediated crowd. As we can see from Figure 1, with the next generation Crowdsourcing techniques, this assumption no longer holds in the same way. With Wearables Crowdsourcing and CrowdSensing, data collection is autonomous of human intervention, passively collected as the individual carries the device in physical space. While it is of course true that an individual can turn-off smartphone sensors -or remove Wearables from the body- such new configurations of IT-mediated crowds would seem to require a re-investigation of intrinsic and extrinsic motivation in IT-mediated crowds, where crowds are either partially or fully comprised of such next generation participation.

In terms of the types of intelligence sought from IT-mediated crowds, to date the research has focused upon tasks, problem-solving, money, and idea-generation (Brabham 2008; 2013, Gerber & Hui 2013, Ipeirotis & Paritosh 2011, Michelucci 2013) as the primary forms of intelligence generated from IT-mediated crowds. Due to the fact that these first generation Crowdsourcing techniques have been tied directly to the desktop computing paradigm, the research has always been limited to crowd contributions in a virtual space, with inputs, generally text, that can only be rendered digitally (Fayard & Metiu 2014). As we can see from Figure 1, with the next generation Crowdsourcing techniques, this assumption no longer holds in the same way. With Situated and Spatial Crowdsourcing, data collection or task-completion occurs solely in a particular physical space, at a particular time, and in the case of Spatial Crowdsourcing, includes material inputs that cannot be digitally rendered, nor transferred. Along similar lines, CrowdSensing inputs, though digitally rendered, are completely tied to the physical environment for their operation, both in the sense that they are tied to the mobility of the individual carrying the device in the physical environment, and in the sense that all inputs are longitudinal measurements of the qualities of the physical environment itself. Similarly, with Wearables Crowdsourcing, we now have the ability to capture digitally rendered longitudinal inputs, solely concerned with the physical state of a particular individual, tied to the mobility of the individual wearing the device in the physical environment.

Altogether, these preliminary findings seem to indicate that there is much new work to be done in the Crowdsourcing domain. However, this treatment of the subject is necessarily limited due to the space considerations of this forum. Many other specific avenues of research inquiry can follow directly from these findings. For example, the Crowdsourcing research illustrates that self-selection is a key antecedent of crowd-member participation (Jeppesen & Lakhani 2010). Further, the cognitive diversity of the individuals comprising IT-mediated crowds, combined with the scale of overall crowd-size, have been put forward as key factors explaining why IT-mediated crowds are effective at generating Collective Intelligence outcomes (Aitamurto 2016). Similarly, first generation Crowdsourcing techniques have been shown to exhibit differences, ranging orders of magnitude, in the size of IT-mediated crowds that they form (Prpić et al 2015b). New research needs to understand whether these fundamental features of Crowdsourcing as we know it, also hold for the new configurations of hardware, software, and people, found in the next generation Crowdsourcing techniques.

**6.0: CONCLUSION**

This brief exploratory work illustrates that next generation Crowdsourcing techniques represent new configurations of hardware, software, and people for organizing IT-mediated crowds. Following a review of the relevant literatures, the work presents a collectively exhaustive and mutually exclusive typology premised directly upon two salient dimensions found in the extant literautre. As a result, this work situates the next generation Crowdsourcing techniques within the extant Crowdsourcing literature and identifies new avenues of research stemming directly from the analysis.